\documentclass[12pt]{article}
\usepackage{fullpage,latexsym,amsmath,amsfonts,amssymb,amsthm,graphicx,wasysym,mathtools}
\usepackage[style = numeric]{biblatex}

\DeclarePairedDelimiter\parens{(}{)}

\newcommand{\ii}[1]{\textit{#1}}
\newcommand{\bb}[1]{\textbf{#1}}

\newcommand{\absolute}[1]{\left | #1 \right |}
\newcommand{\hilbert}{\mathcal{H}}
\newcommand{\statespace}{\mathcal{V}}
\newcommand{\ket}[1]{\left | #1 \right \rangle}
\newcommand{\bra}[1]{\left \langle #1 \right |}
\newcommand{\proj}[1]{\ket{#1} \!\! \bra{#1}}

\newcommand{\amp}[2]{\left \langle #1 | #2 \right \rangle}

\newcommand{\sys}[1]{^{\mbox{\tiny (#1)}}}

\newcommand{\mket}[1]{\left | #1 \right ) }

\newcommand{\modalspace}{\mbox{$\mathcal{V}$}}

\newcommand{\scalarfield}{\mbox{$\mathcal{F}$}}

\newcommand{\vspan}[1]{\left \langle #1 \right \rangle}

\newcommand{\Z}{\mathbb{Z}}

\title{Cloning, deleting and hiding in modal quantum theory}
\author{Phillip Diamond and Benjamin Schumacher}

\begin{document}

\maketitle

\begin{abstract}
 We examine the toy model of modal quantum theory (MQT), an
 analogue of actual quantum theory based on finite fields.
 In particular, we investigate how several essential ``no-go''
 results (for cloning, deleting and hiding processes) work in 
 MQT.  Cloning and deleting are still forbidden in MQT, though the
 details of these results are somewhat different in the new
 context.  However, the information of a modal qubit can 
 be completely hidden in the correlations between two entangled 
 modal qubits.
\end{abstract}

\section{Introduction}

No-go theorems show that a particular process or circumstance 
cannot occur within the context
of a given physical theory.  
Such theorems have proven to be useful tools for understanding 
the essential content of quantum theory, 
and they have played a key role in our 
understanding of quantum foundations and quantum
information theory.  This paper continues a project
of exploring quantum no-go theorems by considering
modal quantum theory (MQT), a mathematical ``toy
model'' of actual quantum theory based on finite
fields.  The structure of MQT lacks the idea of probability,
and the concepts of state, time evolution and 
distinguishability are substantially different from 
actual quantum theory.  

We will find that some 
familiar no-go results have direct analogues 
in MQT, while others do not.  Even in the 
cases where MQT has a parallel no-go result,
the details can be quite dissimilar.  By 
formulating our arguments within MQT, 
we can shed light on the way
no-go theorems work, and even arrive at new ideas 
and arguments applicable to the
quantum theory of our world.

\section{Modal quantum theory}

In modal quantum theory \cite{MQT2012,MQT-almostQT}, the space of states of
a system is a finite-dimensional vector space $\modalspace$ 
over a general field $\scalarfield$, usually taken to be
finite.  There is no inner product, and any non-null vector
$\mket{\psi}$ represents a pure state.  Given a basis
set $\{ \mket{\phi_{i}} \}$ for $\modalspace$, we can write
\begin{equation}
   \mket{\psi}=\sum_{i} \alpha_{i} \mket{\phi_{i}},
\end{equation}
where $\alpha_i\in \mathcal{F}.$ 
The basis vectors $\mket{\phi_{i}}$ are identified with the
possible results of a simple ``basic'' measurement on the system.
But MQT does not predict probabilities for these results.
Instead, MQT only makes the ``modal'' distinction between
possible and impossible results.  For state $\mket{\psi}$, 
the result $\phi_{i}$ is {\em possible} if $\alpha_{i} \neq 0$
and {\em impossible} if $\alpha_{i} = 0$.

The time evolution of an isolated modal quantum system
over a fixed time interval is given by an invertible
linear operator $T$ on $\modalspace$.  The states of 
composite systems are non-null vectors in the tensor
product space $\modalspace\sys{12} = \modalspace\sys{1} \otimes 
\modalspace\sys{2}$.

As in actual quantum theory, we may also consider 
mixed states.  These arise in two distinct situations.
First, we may not have complete information about the
preparation of a system.  Second, the system of
interest may be part of a larger composite system
in an entangled state.  In either case, MQT represents
a mixed state by a subspace $\mathcal{M}$ of $\modalspace$.
Suppose we have a set $\{ \mket{\psi}_{a} \}$ of possible
state vectors for the system.  Then the mixed state is
\begin{equation}
    \mathcal{M} = \vspan{ \{ \mket{\psi}_{a} \} },
\end{equation}
where $\vspan{\cdots}$ denotes the linear span of a set.
A measurement result $\mket{\phi_{i}}$ is possible for 
$\mathcal{M}$ whenever it is possible for some 
$\mket{\psi} \in \mathcal{M}$.

Alternatively, suppose we have a pure entangled state
$\mket{\Psi\sys{12}}$ of a composition of systems 
1 and 2.  If we choose a basis $\mket{\phi_{i}\sys{1}}$
for system 1 states, we can write this as
\begin{equation}
    \mket{\Psi\sys{12}} = \sum_{i} \mket{\phi_{i}\sys{1}}
        \otimes \mket{\psi_{i}\sys{2}}.
\end{equation}
The measurement result $\mket{\phi_{i}\sys{1}}$ for
system 1 is possible when $\mket{\psi_{i}\sys{2}} \neq 0$, 
and if this result does occur then the conditional state
of system 2 is $\mket{\psi_{i}\sys{2}}$.  The mixed state
of system 2 is just $\mathcal{M}\sys{2} = 
\vspan{ \{ \mket{\psi_{i}\sys{2}} \}}$, a subspace
that is independent of the choice of basis $\{ \mket{\phi_{i}\sys{1}} \}$
for $\modalspace\sys{1}$.

To summarize, modal quantum theory retains some of the
structure of actual quantum theory, while discarding other
properties.  The states are elements of a vector space, 
(thus allowing for superpositions) but there is no
inner product on that space.  Measurement results may
be possible or impossible, but no probabilities are entailed.
Time evolution is linear and invertible, but not unitary.
Mixed states arise from incomplete information about 
system preparation or as a subsystem state of an entangled
composite state.  These are described by subspaces rather
than density operators.

The concepts of actual quantum theory and modal quantum
theory are compared in Table~\ref{tab-AQTMQT}.

\begin{table}[t]
\begin{center}
    \begin{tabular}{l|c|c}
        &  AQT  &  MQT  \\[2ex] \hline
        States 
            &  $\ket{\psi} = \sum_{i} c_{i} \ket{\phi_{i}}$, $c_{i} \in \mathbb{C}$
            &  $\mket{\psi} = \sum_{i} \alpha_{i} \mket{\phi_{i}}$, $\alpha_i \in \scalarfield$\\[2ex]
            \hline
        Bases
            &  $\ket{\phi_{i}}$ orthonormal 
            &  $\ket{\phi_{i}}$ linearly independent \\[2ex]
            \hline
        Amplitudes  
            &  $P(\phi_{i}) = \absolute{c_{i}}^2$
            &  $\begin{array}{l} \alpha_{i} \neq 0 \Rightarrow \phi_{i} \mbox{ possible}\\
                \alpha_{i} = 0 \Rightarrow \phi_{i} \mbox{ impossible} \end{array}$ \\[2ex]
                \hline
        Normalization
            & $\amp{\psi}{\psi} = \sum_{i} \absolute{c_{i}}^2 = 1$ 
            & $\mket{\psi} \neq 0$ (at least one $\alpha_{i} \neq 0$) \\[2ex]
            \hline
        Time evolution
            & $\ket{\psi(t)} = U \ket{\psi(0)}$
            & $\mket{\psi(t)} = T \mket{\psi(0)}$  \\[2ex]
            \hline
        Composite states
            & $\ket{\Psi\sys{12}} \in \hilbert\sys{1} \otimes \hilbert\sys{2}$
            & $\mket{\Psi\sys{12}} \in \statespace\sys{1} \otimes \statespace\sys{2}$
            \\[2ex] \hline
        Mixed states
            & $\rho = \sum_{a} p_{a} \proj{\psi_{a}}$
            & $\mathcal{M} = \vspan{ \{ \mket{\psi_{a}} \} }$
    \end{tabular}
\end{center}
\caption{Aspects of actual quantum theory (AQT) and their MQT analogues.
    \label{tab-AQTMQT}}
\end{table}

\section{Three classical operations and their analogs}

The three operations we will consider are suggested 
by operations in classical reversible logic.  Given a
``control'' bit $a$ and a ``target'' bit $b$, the
controlled-not (CNOT) operation is
\begin{equation}
    (a,b) \longrightarrow (a,a \oplus b) ,
\end{equation}
where $\oplus$ is addition mod 2.  Thus,
\begin{equation}
    \begin{array}{ccc}
    (0,0) \rightarrow (0,0) & \quad & (1,0) \rightarrow (1,1) \\
    (0,1) \rightarrow (0,1) & \quad & (1,1) \rightarrow (1,0)
    \end{array} .
\end{equation}
This operation is clearly reversible, since it is its 
own inverse.  We can use the CNOT operation as the basis
for several common tasks.
\begin{description}
    \item[Cloning (or copying).]  If the target bit $b$ is initially 0,
        then its final state will be an exact copy of $a$:
        \begin{equation}
            (a,0) \longrightarrow (a,a) .
        \end{equation}
        Thus, we can make copies of a data $a$ into
        blank memory.
    \item[Deleting (or uncopying).]  If the bits
        are initially equal, then the final state of the
        target bit will be 0:
        \begin{equation}
            (a,a) \longrightarrow (a,0) .
        \end{equation}
        Thus, we can delete one of our two copies of the 
        same information.
    \item[Hiding.]  Suppose the control bit is a cryptographic 
        key $k$ and the target is a plaintext message bit $p$.
        Then
        \begin{equation}
            (k,p) \longrightarrow (k,c)
        \end{equation}
        where $c=k \oplus p$, the encrypted ciphertext.
        Neither $k$ nor $c$ alone provides any information 
        whatsoever about the input plaintext $p$, which
        can only be recovered from the pair $(k,c)$.  In
        other words, the $p$-information is completely
        ``hidden'' in the correlations between $k$ and $c$.
\end{description}
In actual quantum theory, the analogues of these three 
classical operations are forbidden by no-go theorems. 
\begin{description}
    \item [Quantum no-cloning theorem.]  No process can duplicate
        an unknown input quantum state, taking $\ket{\psi}$
        as input and yielding $\ket{\psi, \psi}$ as output.
        \cite{wznocloning,ddnocloning}
    \item  [Quantum no-deleting theorem.]  If a process deletes
        one of a duplicate pair of states---taking $\ket{\psi,\psi}$
        as input and yielding $\ket{\psi,0}$ as output---then an
        isomorphic copy of state $\ket{\psi}$ must be retained in 
        the final state of the auxiliary apparatus.
        \cite{nodeleting2000}
    \item  [Quantum no-hiding theorem.]  The information of an
        unknown qubit state cannot be stored entirely in the
        correlation between two qubits.  \cite{nohiding2007}
\end{description}
In modal quantum theory, as we will see, the situation is more
complicated.



\section{Distinguishability, Cloning, and Deleting}

In AQT, two unknown states are \textit{distinguishable} if they are orthogonal, so that there exists a measurement that always identifies one state from the other. 
If $\ket{\psi}$ and $\ket{\phi}$ are distinct but not orthogonal, they cannot be perfectly distinguished. Many copies of the same states \begin{equation}
    \underbrace{\ket{\psi}\otimes\ket{\psi}\cdots\ket{\psi}}_\text{$n$ times}\text{ and } \underbrace{\ket{\phi}\otimes\ket{\phi}\otimes\cdots\otimes\ket{\phi}}_\text{$n$ times},
\end{equation} 
may be approximately distinguished, since their inner product \begin{equation}
    \big(\bra{\psi}\otimes\bra{\psi}\otimes\cdots\otimes\bra{\psi}\big) \big(\ket{\phi}\otimes\ket{\phi}\otimes\cdots\otimes\ket{\phi}\big)= \amp{\psi}{\phi}^n
\end{equation} 
approaches 0 as $n\to \infty$. Nevertheless, perfect distinguishability is not possible for finite $n$.

In MQT, two modal states $\mket{\psi}$ and $\mket{\phi}$ can be distinguished 
as long as $\mket{\phi}$ is not parallel to $\mket{\psi}$. 
In general, linearly independent sets of states in MQT can be distinguished, 
and linearly dependent sets cannot. What about $n$-copy states? 
We begin with a lemma.

\bb{Lemma.} Given a set of $d+1$ distinct modal states spanning a $d$-dimensional subspace, the set of $d+1$ two-copy states is distinguishable.
\begin{proof}
    Let $\mathcal{F}$ be the finite field of MQT state coefficients. Assume $S = \{\mket{\psi_1},\mket{\psi_2},...,\mket{\psi_d},\mket{\sigma}\}$ is a set of $d+1$ vectors spanning a $d$-dimensional subspace, where 
    \begin{equation}
        \mket{\sigma} =\sum_{k=1}^d\sigma_k\mket{\psi_k}
    \end{equation} 
    for some set $\{\sigma_k\}\in \mathcal{F}$ such that at least two  $\sigma_k$ are nonzero. Assume $\sigma_1, \sigma_2\neq 0$.  Now consider the same set, but with an additional copy of each state: \begin{equation}
        S^{(2)}=\{\mket{\psi_1,\psi_1},\mket{\psi_2,\psi_2},...\,,\mket{\psi_d,\psi_d},\mket{\sigma,\sigma}\}.
    \end{equation}
    Since the original $\mket{\psi_{k}}$'s are linearly independent, the joint states $\mket{\psi_j,\psi_k}$ are also linearly independent.
    The state $\mket{\sigma,\sigma}$ is evidently not a linear combination of the $\mket{\psi_k,\psi_k}$ states since it includes terms such as $\sigma_1\sigma_2\mket{\psi_1,\psi_2}$.  Hence, $S^{(2)}$ is a linearly independent---and thus distinguishable---set of MQT states.
\end{proof}

Using this lemma inductively, we find that any finite, linearly dependent set of $d$ distinct states can eventually be distinguished given a sufficient (but finite) number of state copies. It suffices to double the number of state copies repeatedly until every (previously) dependent state can be distinguished. 

This result has implications for the no-cloning and no-deleting theorems in MQT. For any system in MQT with finite $\mathcal{F}$ and finite dimension $\mathcal{V}$, there are finitely many distinct states. From the generalized lemma, those states are distinguishable given sufficiently many copies. Let $S$ be a set of states, and let $N$ denote the number of state copies sufficient to distinguish every state in $S$. Since the $N$-copy states are distinguishable, we can define an invertible linear map $T$ which clones from $N\to N+1$ state copies and deletes (via $T^{-1}$) from $N+1\to N$ state copies. Thus both cloning and deleting are possible with sufficient input state copies. 

Now let $M$ denote the minimum number of state copies necessary to distinguish $S$. Cloning is not possible when the number of input state copies is less than $M$ because the set of inputs will be linearly dependent. This can be shown using an argument analogous to Wootters and Zurek \cite{wznocloning}. Any machine that clones linearly independent states will fail to clone their superposition. 

Deleting is also not possible if the number of input state copies is less than $M$, as the input states are linearly dependent. Like cloning, if a machine  deletes all linearly independent states in $S$, it will fail to delete linearly dependent states and will instead map them linearly into the machine state. This is analogous to the argument of Braunstein and Pati \cite{nodeleting2000}. Interestingly, a form of deleting is still possible when deleting from $M\to M-1$ state copies, in which the set of input states are distinguishable but the output states are not. It is not possible to distinguish the outputs of this  deleting process; but since the $M$-copy inputs are distinguishable, a deleting machine could store a classical identifier of the input state, enabling the output state to be reconstructed.

\section{Hiding}

In a hypothetical hiding process, the unknown state of a single qubit is mapped into the 4-dimensional state space of a pair of qubits.  The quantum information is hidden if the output states of the individual qubits are each independent of the input.   This can only be true if the output states are all entangled and reduce to identical mixed states on the individual qubits.  (If the output
states were product states, of course, at least one of the subsystem states would have to depend on the input state.)  This condition is necessary in both actual and modal quantum theory.

We begin therefore by presenting an alternate proof of the no-hiding 
theorem for qubits in AQT by proving that every 2D subspace of the 
4D Hilbert space of a qubit pair must contain at least one product state.  





\begin{proof}
We define an arbitrary transformation from the 2D subspace of one qubit to the 4D subspace of two qubits as follows: 
\begin{eqnarray}
\notag\ket{0} & \to & \ket{\Psi_0} =  \sqrt{\lambda}\ket{0,0}+\sqrt{1-\lambda}\ket{1,1} \\
 \ket{1} & \to & \ket{\Psi_1} = \ket{0}\otimes \big(C_{00}\ket{0}+C_{01}\ket{1}\big)+\ket{1}\otimes\big(C_{10}\ket{0}+C_{11}\ket{1}\big),
\end{eqnarray}
where $\ket{0}$ and $\ket{1}$ are Schmidt basis vectors for $\ket{\Psi_0}$, and the coefficients $C_{ij}\in \mathbb{C}$. If $\ket{\Psi_0}$ is entangled, then $0<\lambda<1$. The generic composite state $\ket{\Psi_1}$ may or may not be entangled.

Applying the transformation on an arbitrary superposition of input states yields 
\begin{eqnarray}\label{generictransformation}
\alpha\ket{\Psi_0}+\beta\ket{\Psi_1} & = & 
\ket{0}\otimes\big((\alpha\sqrt{\lambda}+\beta C_{00}\big)\ket{0}+\beta C_{01}\ket{1}\big) \nonumber \\
 &  &  \qquad +\ket{1}\otimes\big(\beta C_{10}\ket{0}+(\beta C_{11}+\alpha\sqrt{1-\lambda})\ket{1}\big).
\end{eqnarray}
Under what circumstances is Equation \ref{generictransformation} a product state? This will happen if the conditional states of the second subsystem are complex scalar multiples of each other. Explicitly,
\begin{align}
\alpha\sqrt{\lambda}+\beta C_{00}&= \beta C_{10}z\\ \beta C_{01}&=\big(\beta C_{11}+\alpha\sqrt{1-\lambda}\big)z,
\end{align}
for some scalar $z\in \mathbb{C}$.
Solving this system of equations yields a quadratic polynomial in $z$:
\begin{equation}
    \parens*{C_{10}\sqrt{1-\lambda}}z^2+\parens*{C_{11}\sqrt{\lambda}-C_{00}\sqrt{1-\lambda}}z- C_{01}\sqrt{\lambda} = 0
\end{equation}
Since $z\in \mathbb{C}$, this quadratic is solvable regardless of $\lambda$ and $C_{ij}$. So given any $1\to 2$ qubit transformation (any $\lambda$ and normalized set of coefficients $C_{ij}$), we can use the $z$ satisfying the system of equations to solve for $\alpha$ and $\beta$: first set $\beta=1$, then solve the system for $\alpha$ and re-normalize. The resulting $\alpha\ket{\Psi_0}+\beta\ket{\Psi_1}$ and will be a product state.
Thus any transformation from the 2D subspace of one qubit to the 4D subspace of two qubits will contain a product state, violating the requirements for hiding. 
\end{proof}

This algebraic proof, based on linearity but not unitarity, was suggested by analysis of the analogous problem in MQT. In MQT, the system of equations that determines whether hiding is possible yields the quadratic $$C_{10}k^2+(C_{11}-C_{00})k-C_{01} = 0$$ where $k\in\mathcal{F}$, some finite field, and the coefficients $C_{ij}$ are defined as in the preceding proof (note the absence of the Schmidt parameter $\lambda).$ The possibility of hiding is dependent on whether this arbitrary quadratic is ever irreducible over $\mathcal{F}$. In fact, irreducible quadratics exist over every finite field, so we conclude that hiding is \ii{always} possible in MQT.

Further, we can use an irreducible quadratic over a finite field to construct a hiding transformation in MQT. Consider MQT with $\mathcal{F} = \Z_3$. The polynomial $x^2+x+1$ is irreducible over $\Z_3$, so we define $C_{10}=1$, $1 = C_{11}-C_{00},$ and $-C_{01} = 1$. We assume $C_{00} = 1$. Plugging the $C$'s into the same transformation described in the AQT proof (without normalization) yields $$\mket{0}\rightarrow \mket{0,0}+\mket{1,1}$$ and $$\mket{1}\rightarrow \mket{0}\otimes \big(\mket{0}- \mket{1}\big)+\mket{1}\otimes \big(\mket{0}+2\mket{1}\big).$$
Taking an arbitrary superposition of the transformation acting on $\mket{0}$ and $\mket{1}$, then requiring the result to simplify to a product state yields the system of equations 
\begin{align*}
    a+b &= kb\\
     -b &= ka+2kb ,
\end{align*}
leading to the irreducible quadratic
$$k^2 + k +1 = 0 .$$
Thus, the transformation
\begin{align*}
    &\mket{0}\rightarrow \mket{0,0}+\mket{1,1}\\&\mket{1}\rightarrow \mket{0}\otimes \big(\mket{0}- \mket{1}\big)+\mket{1}\otimes \big(\mket{0}+2\mket{1}\big) 
\end{align*}
Will hide any input modal qubit, since there is no $k$ satisfying the system of equations that determines whether the transformation will simplify into a product state.

\section{Remarks}

To summarize, the MQT toy model has both no-cloning and no-deleting theorems.
However, if we begin with sufficiently many copies of the input states, both
cloning and deleting are possible, since these $n$-copy states are perfectly
distinguishable.

In contrast to actual quantum theory, on the other hand, 
in MQT we can always hide the information
in a modal qubit in the correlations between two entangled qubits.
The possibility of qubit hiding in MQT arises from the existence of 
irreducible quadratic polynomials in a finite field $\scalarfield$.
Yet there is more to be said.  A key property of MQT is that all entangled 
states of a pair of qubits have the same subsystem states:  $\modalspace$,
the entire two-dimensional state space.  In actual quantum theory, 
entangled qubits yield subsystem states described by density operators.
Even if two density operators are both supported on the entire state
space, they may differ in the probabilities of different measurement
effects.  Since modal quantum theory has no probabilities, all mixed
qubit states are identical.

In a separate paper \cite{MQTbroadcast}, 
we examine another MQT analogue to a basic
no-go theorem of actual quantum theory:  the no-broadcasting theorem,
the mixed state generalization of the no-cloning theorem.  These 
and other results help illuminate the structure of quantum theory
by showing how its basic concepts change---or remain the same---in
the simpler ``foil'' theory of MQT.

We would like to thank Michael D. Westmoreland for many useful 
conversations and suggestions during our work, and we gratefully
acknowledge the support of the Kenyon Summer Science Scholar program
in the summer of 2022.

\printbibliography

\end{document}